# Time-Domain Signatures of Distinct Correlated Insulators in a Moiré Superlattice

Eric A. Arsenault[1], Yiliu Li[1], Birui Yang[2], Takashi Taniguchi[3], Kenji Watanabe[4], James C. Hone[5], Cory R. Dean[2], Xiaodong Xu[6], and X.-Y. Zhu[1,*]

[1] Department of Chemistry, Columbia University, New York, NY 10027, USA
[2] Department of Physics, Columbia University, New York, NY 10027, USA
[3] Research Center for Materials Nanoarchitectonics, National Institute for Materials Science, 1-1 Namiki, Tsukuba 305-0044, Japan
[4] Research Center for Electronic and Optical Materials, National Institute for Materials Science, 1-1 Namiki, Tsukuba 305-0044, Japan
[5] Department of Mechanical Engineering, Columbia University, New York, NY 10027, USA
[6] Department of Physics, University of Washington, Seattle, WA 98195, USA

**ABSTRACT.** Among expanding discoveries of quantum phases in moiré superlattices, correlated insulators stand out as both the most stable and most commonly observed. Despite the central importance of these states in moiré physics, little is known about their underlying nature. Here, we use pump-probe spectroscopy to show distinct time-domain signatures of correlated insulators at fillings of one ($v = -1$) and two ($v = -2$) holes per moiré unit cell in the angle-aligned WSe$_2$/WS$_2$ system. Following photo-doping, we find that the disordering time of the $v = -1$ state is independent of excitation density ($n_\text{ex}$), as expected from the characteristic phonon response time associated with a polaronic state. In contrast, the disordering time of the $v = -2$ state scales with $1/\sqrt{n_\text{ex}}$, in agreement with plasmonic screening from free holons and doublons. These states display disparate reordering behavior dominated either by first order ($v = -1$) or second order ($v = -2$) recombination, suggesting the presence of Hubbard excitons and free carrier-like holons/doublons, respectively. Our work delineates the roles of electron-phonon (e-ph) versus electron-electron (e-e) interactions in correlated insulators on the moiré landscape and establishes non-equilibrium responses as mechanistic signatures for distinguishing and discovering quantum phases.

---

[*] To whom correspondence should be addressed. E-mail: xyzhu@columbia.edu.

**Introduction**

Moiré materials offer unprecedented and highly tunable platforms for understanding the emergence of quantum phases of matter and for exploring future device applications. Among the quantum phases realized in these systems, the correlated insulators stand out as the most commonly observed[1–15]. Other quantum phases, including mixed bosonic-fermionic states[16–18], superconductors[19,20], and various quantum Hall states[21–28], are often directly related to or appear in the vicinity of correlated insulators. Despite the obvious importance of correlated insulators in moiré systems, an understanding of the hierarchy of stabilizing/destabilizing interactions remains largely limited to a reliance on simple models, such as the Hubbard Hamiltonian[29,30]. Establishing how correlated insulators differ mechanistically is essential to understanding their properties, such as varying critical temperatures ($T_c$), and connections to other quantum phases. We focus on the most robust quantum states discovered to date at moiré interfaces—correlated insulators at integer filling factors of moiré superlattices in transition metal dichalcogenide (TMD) heterobilayers[2–15]. We choose the angle-aligned $WSe_2/WS_2$ moiré system, where the $v = -1$ and $v = -2$ correlated insulator states exhibit high $T_c$ of ~150 K and ~80 K, respectively[2,3,6,8,11]. Recently, we suggested that the observed high $T_c$ is, in addition to many-body e-e interactions, related to polaronic stabilization due to e-ph interactions[31].

In the angle-aligned $WSe_2/WS_2$ moiré system studied here, there are two potential wells at high symmetry points in each moiré unit cell, leading to the formation of two narrow moiré bands[32,33]. At half filling ($v = -1$) of the first moiré band (lower energy), many-body correlation results in gap opening near the top of the valence band and the formation of an upper and a lower Hubbard band (UHB and LHB, respectively). The second moiré band is believed to be located within the Hubbard gap of the first moiré band and the $v = -1$ correlated insulator is properly called a charge-transfer (CT) insulator[32] instead of a Mott insulator[34]. At $v = -2$, the inter-band Coulomb energy ($U'$, rather than the intra-band Coulomb energy, $U$) favors the half filling of both bands and the result is a two-band Mott insulator[16] rather than a trivial band insulator[6]. To understand these correlated insulators, we take a time-domain approach[31] in which a pump pulse creates holons and doublons across the correlated gap and a probe pulse detects the disordering dynamics[35] via exciton sensing of the dielectric environment[2,3,6,8]. Such an approach directly accesses the timescales relevant to electronic or atomic motions that are not accessible in steady-state spectroscopy or transport measurements. In particular, disordering of a correlated state occurs on a timescale of the characteristic system response, linked to either electron hopping or the oscillation period of a relevant phonon mode, depending on whether e-e or e-ph interactions play the dominate role in stabilizing the state[36]. Here, we apply such an approach coupled with fluence-dependent measurements. Through this lens, we map the mechanistic regimes governing correlated insulator stability/instability in moiré systems as a function of photoexcitation density, $n_{ex}$.

**Steady-State and Transient Correlated State Sensing**

Fig. 1a illustrates the WSe$_2$/WS$_2$ heterostructure (AB stacked, $\theta = 60\pm1°$) featuring top and bottom gates, each consisting of a hexagonal boron nitride (h-BN) dielectric spacer and a few-layer graphene (Gr) electrode. The dual-gate structure allows either hole ($v < 0$) or electron ($v > 0$) doping from an applied gate voltage (V$_g$), while the electric field at the moiré interface can be maintained at zero. Fig. 1b illustrates the static reflectance spectrum of the device as a function of V$_g$ and moiré filling factor $v$ (defined here as charge filling per moiré unit cell). In the reflectance measurement, the excitonic oscillator strength, sensitive to changes in the dielectric environment, is an effective probe of correlated state formation. We choose the lowest energy moiré exciton of WSe$_2$ as our dielectric sensor for its high oscillator strength and for its sensitivity to hole doping, as the holes reside exclusively in the WSe$_2$ layer due to the type-II band alignment at the WSe$_2$/WS$_2$ interface[2,3]. At $v = \pm1, \pm2$, we observe an increased exciton signal as the effective dielectric constant decreases upon the formation of correlated insulator states[2,3,6,8]. Further device characterization can be found in Fig. S1.

By introducing pulsed laser excitation (and detection), we can further perform transient reflectance measurements on the same device architecture[31]. This allows us to obtain both time-resolved and gate-dependent reflectance spectra (see Fig. S2-S3). For excitation, we employ a pump with photon energy (1.55 eV) below the optical gaps of WSe$_2$ and WS$_2$ to uniquely target the correlated states. This avoids pump-induced interlayer exciton formation which would result in long-lived responses on the nanosecond time scale[33]. The absorption of a pump photon excites a hole from the LHB of the correlated insulator deep into the valence band(s). The excited holes relax within this continuous manifold on ultrafast time scales, typically of the order of 10s-100s femtoseconds due to scattering with other carriers and optical phonons[37], towards the upper unoccupied band(s) (the UHB or the CT band)[32]. The result is the formation of holon-doublon pairs across the correlation gap (see Fig. S4 for schematic). The presence of holons and doublons leads to disordering of the correlated insulator on characteristic timescales determined by e-e or e-ph interactions[36]. Recovery of the charge order reaches the so-called "bottleneck" on a timescale determined by recombination of holons and doublons across the gap[35]. The timescale of disordering is captured by the transient increase in the effective dielectric constant, monitored through the exciton sensor covered by a probe pulse centered at 1.65 eV (spanning the lowest energy moiré exciton of $WSe_2$). Correlated state recovery is tracked through the subsequent reduction in the effective dielectric constant as the original insulator behavior reemerges.

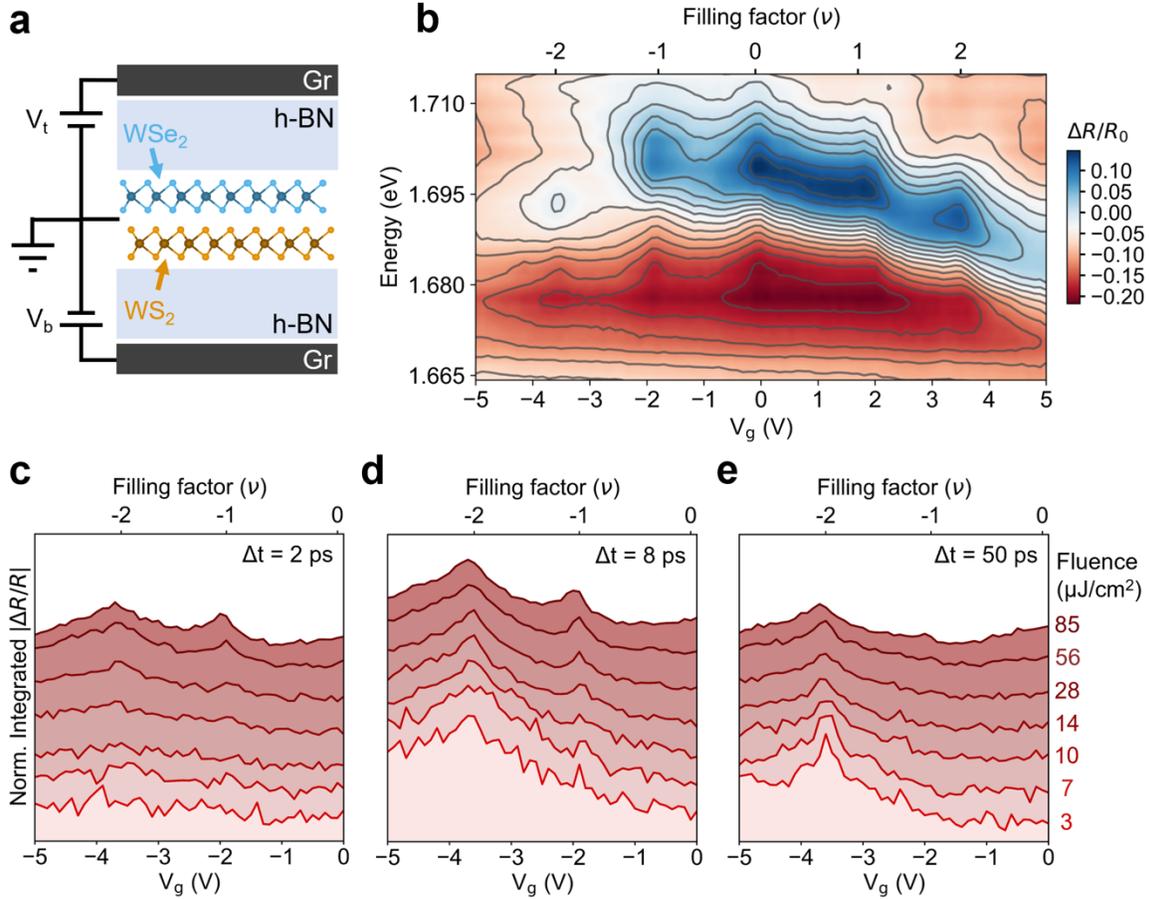

**Figure 1. Steady-state and transient correlated state detection via exciton sensing. a** Schematic of device architecture where the $WSe_2/WS_2$ heterobilayer is symmetrically encapsulated by top and bottom gates ($V_t$ and $V_b$, respectively) consisting of hexagonal boron nitride (h-BN) and few-layer graphene (Gr). **b** Gated ($V_g = V_t = V_b$) steady-state reflectance spectrum of the lowest energy moiré exciton of $WSe_2$. Here, $\Delta R = R - R_0$ where $R$ is the reflected signal from the bilayer and dual gate region while $R_0$ is the reflected signal from a region with gates and no bilayer. **c-e** Gated ($V_g = V_t = V_b$) transient reflectance response as a function of pump fluence at $\Delta t = 2, 8, 50$ ps, respectively. Shown is the integrated response centralized around the lowest energy $WSe_2$ moiré exciton sensor. The different fluence-dependent responses have been offset and multiplied by -1 for clarity. Associated fluence labels are shown on the side of panel e and are consistently colored throughout the other panels. Here, $\Delta R = R_{on} - R_{off}$ where the subscript indicates either pump-on or -off spectra. See Methods for further details. All data were collected at 11 K.

Fig. 1c-1e show transient reflectance spectra integrated around the lowest energy moiré transition of WSe$_2$ as functions of both applied gate voltage, $V_g$, and pump-probe time delay, $\Delta t$, at varying pump fluences. The pump fluence is increased until saturating behavior is reached in the transient response (Fig. S5). In agreement with the static reflectance spectra, correlated state responses at $v = -1, -2$ are observed in the time-resolved spectra, with the maximum response (disordering) in a few ps and recovery (reordering) in tens of ps. With increasing fluence, the magnitude of the correlated state response also increases. Intriguingly, the response of the $v = -2$ state to pump-induced disordering occurs in a broader $V_g$ range than the $v = -1$ state does (Fig. 1c-1e), in contrast to the steady-state measurement (Fig. 1b). This may suggest that the transient measurements are particularly sensitive to an expanse of otherwise hidden states with varying charge ordering around $v = -2$, as predicted for lower filling factors[38,39]. In the present case, the magnitude of the transient response around $v = -2$ indicates a significant change in the dielectric environment and therefore a substantial non-equilibrium photoinduced response. The states in this region also appear to have varying recovery timescales (Fig. 1e) which can be seen as a narrowing in the transient response along the $V_g$ axis at later $\Delta t$, particularly at lower pump fluence. In the following, we focus on the disordering and reordering dynamics at $v = -1, -2$.

**Distinct Temporal Responses of Correlated Insulator States**

Fig. 2a-2b show the transient reflectance responses of the $v = -1, -2$ states for the indicated pump fluences ($f = 3$-$85$ µJ/cm$^2$). Following the coherent artifact arising from pump-probe overlap at $\Delta t = 0$ ps, the non-equilibrium response of the correlated states is initially related to disordering of the equilibrium charge configuration, as evidenced by an increase in the effective dielectric constant detected through bleaching of the exciton transition. In each case, the disordering reaches a maximum magnitude after a few ps and subsequently recovers. The black solid curve in each case is a fit consisting of a single exponential decay (disordering) and either a single exponential first order ($v = -1$) or a second order ($v = -2$) recovery (reordering); further fit details can be found in the Methods section (fit residuals shown in Fig. S6). The chosen functional form for the recovery dynamics will be explicitly discussed further below. Finally, an additional offset is also included to account for the much longer timescale (>>100 ps) recovery dynamics. This offset is negligible at low excitation fluence but becomes significant at $f > 14$ µJ/cm$^2$. As detailed in control experiments on the undoped charge neutral state ($v = 0$, Fig. S8) the long-lived offset can be attributed to hole injection into the WSe$_2$/WS$_2$ moiré structure from thermionic emission of hot

electrons in the unavoidably photo-excited Gr electrodes at sufficiently high excitation densities (Fig. S9)[40,41]. Concomitant with the dynamic responses from disordering and recovery are temporal oscillations of interlayer phonon modes which have been discussed elsewhere[42]. These modes are coherent phonon wavepackets launched from strain fields induced by the pump at the Gr electrodes. These wavepackets propagate through the h-BN prior to impinging on the heterobilayer with an apparent delay as determined by the phonon group velocity.

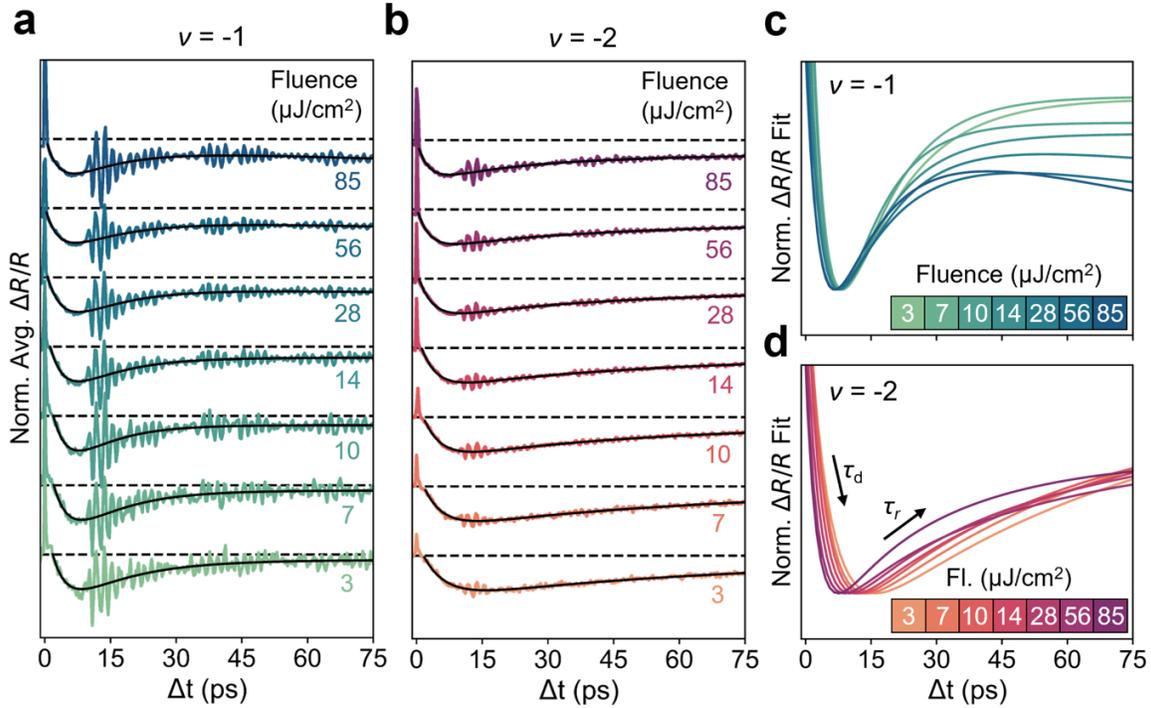

**Figure 2. Non-equilibrium correlated insulator response. a-b** Fluence-dependent temporal response of the $v = -1, -2$ states, respectively. The time traces were averaged in a narrow window about the maximum of the lowest energy $WSe_2$ moiré exciton sensor. Each fluence-dependent transient has been normalized for clear comparison of the dynamical response (see Fig. S7 for unnormalized data where the overall signal trend is preserved). Data are shown in shaded colors while the fits used for time constant extraction are shown in black. See the Methods section for further details. Throughout the panels, increasing pump fluence is shown from light to dark shades. All fluence-dependent data are colored consistently for each of the filling factors throughout the figure. All data were collected at 11 K. **c-d** To more easily and directly visualize the trends in the fluence-dependent dynamics (without the substantial interlayer phonon response), the normalized fits from panels a and b are shown in panels c and d, respectively. Arrows indicating the time constants associated with the disordering dynamics ($\tau_d$) and recovery dynamics ($\tau_r$) are labeled in panel d.

To more clearly view the disordering and recovery dynamics without interference from the coherent phonon oscillations, we replot the normalized fits independently from the data in Figs. 2c-2d for $v = -1, -2$, respectively. The disordering and recover dynamics of the $v = -1$ state are found to be independent of fluence in the entire investigated range ($f$ = 3-85 µJ/cm$^2$). In contrast, the $v$ = -2 state exhibits distinct behavior—both the disordering and the recovery accelerate with increasing fluence. To quantitatively compare the two correlated insulator states, we plot in Fig. 3a and Fig. 3b the disordering time ($\tau_d$) for the at $v$ = -1 and -2 states, respectively, as functions of fluence.

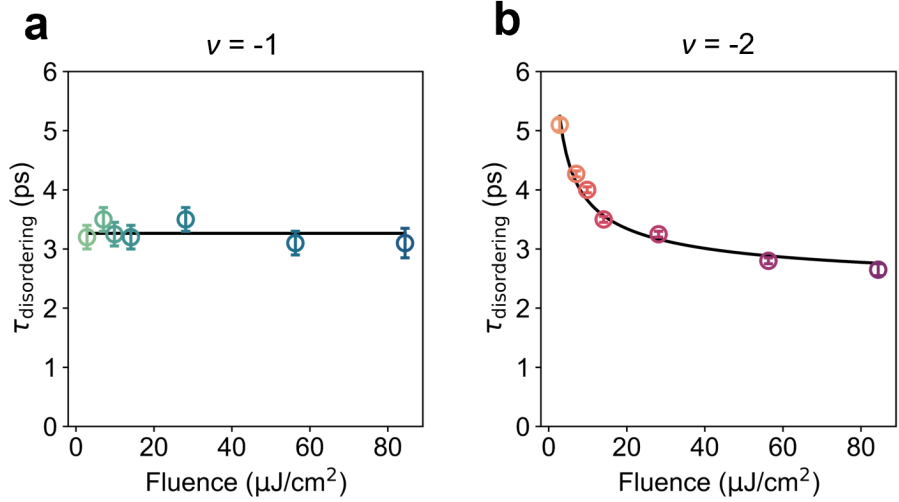

**Figure 3. Disordering in the polaronic versus plasmonic limits. a** Disordering time constant ($\tau_d$) for $v$ = -1 as a function of fluence. The linear fit trend line is shown in black. **b** Disordering time constant ($\tau_d$) for $v$ = -2 as a function of fluence. Shown in black is the fit of the form $\propto 1/\sqrt{n_{ex}} + \tau_{const}$. Throughout the figure, all error bars indicate 99% confidence intervals. Large circle outlines have been added to aid with data visualization. The fluence-dependent shaded colors for the $v$ = -1, -2 states are consistent with the data shown elsewhere in the text. See the Methods section for further details.

For the $v$ = -1 state, disordering occurs with a time constant $\tau_d$ = 3.3 ± 0.2 ps, independent of the photoexcited carrier density, $n_{ex}$, of the correlated insulator (up to saturation of the observed transient response in the investigated pump fluence range). This result precludes a pure electronically-driven disordering mechanism, the rate (=1/$\tau_d$) of which should increase with $n_{ex}$ due to carrier screening of the many-body electronic interactions. Instead, the constant $\tau_d$ suggests a time scale characteristic of e-ph interactions and rate-limited by the oscillation period of the relevant phonon mode(s), often described as the amplitude mode in charge density wave (CDW) systems[36,43]. Our initial pump-probe measurements on the $v$ = -1 correlated insulator state in the WSe$_2$/WS$_2$ moiré superlattice revealed thermal activation behavior in the disordering process, suggesting the polaronic nature of the correlated insulator, in agreement with an *ab initio* calculation[31]. The independence of disordering rate on $n_{ex}$ observed here not only confirms the polaronic interpretation, but also reveals the dominance of e-ph interactions in stabilizing the charge order in the $v$ = -1 correlated insulator. Such a polaronic and ordered correlated state is similar to a CDW state[44]; however, unambiguous delineation between Mott insulators versus CDW ordering remains a subject of discussion[45].

The observed $\tau_d$ can be understood as relating to the oscillation period (typically taken as either a half or quarter period) of phonon mode(s) involved, as shown in the related CDW systems[36,43]. The corresponding mode frequency is consistent with the fact that the moiré deformation potential likely involves acoustic phonons accessible via the momentum range defined by the mini-Brillouin zone of the moiré superlattice[31]. Interestingly, in the fit residuals, a low frequency oscillation attributable to this phonon mode is potentially present at $\Delta t < 10$ ps; however, this remains speculative as this feature is obscured by the interlayer modes after only a single period (Fig. S6). Further suggestive, this behavior is noticeably absent in the control $v$ = 0 data (Fig. S8).

In stark contrast to the behavior of the $v$ = -1 state, the disordering time, $\tau_d$, for the $v$ = -2 state displays a strong dependence on fluence (Fig. 3b). This is unexplainable by only considering e-ph interactions and instead suggests the importance of many-body electronic interactions. In particular, we find that the disordering time can be well-described by the functional form $\tau_d \propto 1/\sqrt{n_{ex}} + \tau_{const}$ (solid black curve in Fig. 3b). The square root dependence on doping density is well-known for the 2D plasmon frequency[36,43]. In other words, disordering as a result of photoexcited carrier screening occurs on a characteristic timescale inversely proportional to the plasmon frequency, $\omega_p$[36], as reported for the ultrafast collapse of order in the excitonic insulator 1T-TiSe$_2$[43]. We conclude that disordering of the $v$ = -2 correlated insulator at the WSe$_2$/WS$_2$ moiré interface observed here is determined by the electronic response, namely screening by photodoped holons and doublons. Our previous measurement on temperature-dependent disordering of the $v$ = -2 state shows a thermal activation energy lower than that of the $v$ = -1 state [31]. The current result, in combination with our previous finding, reveals that plasmonic screening overwhelms polaronic stabilization in the disordering of the $v$ = -2 correlated insulator. We note that, while the observed $\tau_d$ is dominated by screening as shown by the $1/\sqrt{n_{ex}}$ dependence, there seems to be a fundamental limit to $\tau_d$ as captured by the offset, $\tau_{const} = 2.2 \pm 0.1$ ps. The exact origin of this offset is not known, but we speculate it may be associated with either additional e-ph interactions[31] or a deviation from the pure 2D plasmon response in the moiré superlattice[46]. Papaj and Lewandowski recently theoretically proposed that the coupling of charge order to the collective plasmon response can result in plasmon band folding and gap opening, leading to an upper bound in the plasmon frequency[46]. These issues deserve further investigations.

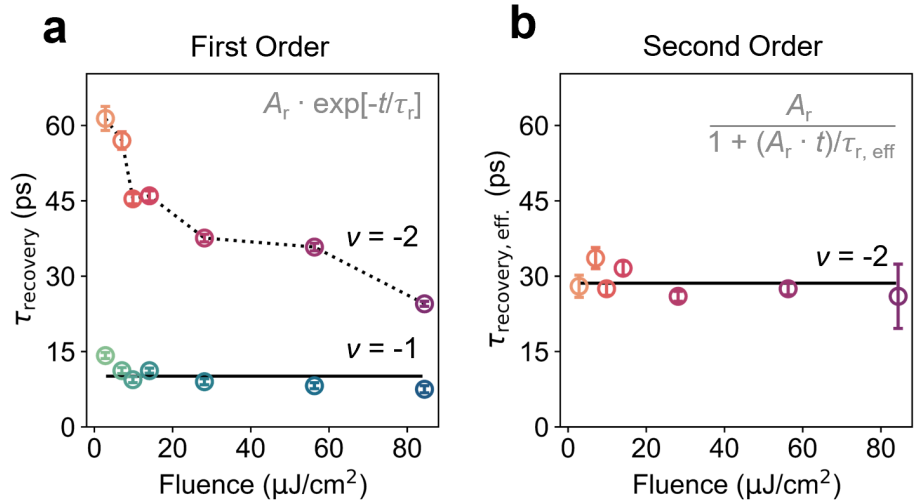

**Figure 4. Charge reordering of distinct correlated insulators. a** Reordering time constant ($\tau_r$) for $v = -1$ and $v = -2$ as a function of fluence. Here, $\tau_r$ was extracted based on considering the recovery as a first order process (mono-exponential functional form, inset in panel a). The black line for $v = -1$ shows the linear fit trend line for $\tau_r$ as a function of fluence. The dotted black line for $v = -2$ shows the linear interpolation between data points to highlight the data trend. **b** Effective reordering time constant ($\tau_{r, eff}$) for $v = -2$ as a function of fluence. Here, $\tau_{r, eff}$ was extracted based on considering the recovery for $v = -2$ as a second order process (inset in panel b). The black line shows the linear fit for $\tau_{r, eff}$ as a function of fluence. Throughout the figure, all error bars indicate 99% confidence intervals. Large circle outlines have been added to aid with data visualization. The fluence-dependent shaded colors for the $v = -1, -2$ states are consistent with the data shown elsewhere in the text. See the Methods section for further details and Fig. S10 for plots of $k_r$ (= $1/\tau_r$).

The fundamental distinction between the $v = -2$ and $v = -1$ insulator states can be understood from the difference in charge occupation of each moiré unit cell. In the $v = -2$ state, there are two holes (on two different sites) within each moiré unit, giving rise to the Coulomb repulsion energy $U'$. For comparison, $U'$ is absent for the $v = -1$ state where the second site remains unoccupied. The magnitude of $U'$ (on the order of 50~100 meV)[16,32] dominates over the polaron stabilization energy (~13 meV estimated from thermal activation)[31]. As a result, photodoping of the two-band correlated insulator, $v = -2$, leads to a free carrier-like plasmonic response. In contrast, the stronger polaronic localization for the $v = -1$ state diminishes the collective plasmon character for holons and doublons.

The fundamental distinction between $v = -1$ and $v = -2$ is also reflected in the recovery dynamics, which are determined by holon-doublon recombination across the correlation gap[35]. Fig. 4a shows the reordering time constants obtained from single exponential fits to the recovery component of the data in Fig. 2. For the $v = -1$ state, the reordering is captured as a first order process implied in the mono-exponential fit and verified by the approximate independence of recovery time constants $\tau_r$ on $n_{ex}$. Such a first-order kinetic process reveals that recombination of the perturbed $v = -1$ state derives from a "single body", i.e., a local holon-doublon pair called a Hubbard exciton[47]. This local holon-doublon pair is fortified by polaronic stabilization, which inhibits their separation into a free holon and a free doublon.

However, for $v = -2$, the single-exponential recovery time constant $\tau_r$ decreases with $n_{ex}$, indicating the failure of the first order kinetic model and that the recovery process is of higher order in $n_{ex}$ (see Fig. S10a). We find that the recovery for $v = -2$ is second order, i.e., the recovery rate is proportional to $n_{ex}^2$. To fit the recovery to a second order kinetic scheme, we use the functional form $A_r/(1+A_r \cdot t/\tau_{r,\text{eff}})$, where $A_r$ is a constant proportional to $n_{ex}^{(0)}$, the initial concentration of photo-excitation as controlled via the pump fluence. Fig. 4b shows the resulting effective second order recovery time constant, $\tau_{r,\text{eff}}$, which is correctly $n_{ex}$-independent (see Fig. S10b). The apparent second order process supports the two-body recombination kinetics of a holon and a doublon, in excellent agreement with the free-carrier nature of photo-doping of the $v = -2$ correlated insulator state.

In summary, we find distinct disordering and reordering mechanisms for the $v = -1$ and -2 states in the WSe$_2$/WS$_2$ moiré superlattice. The non-equilibrium response of the $v = -1$ state reveals its dominant polaronic nature, leading to disordering determined by a characteristic phonon time and re-ordering by the first-order recombination of a localized holon-doublon pair. In contrast, in the $v = -2$ state, e-e interactions dominate over e-ph interactions. As a result, photodoping of the $v = -2$ state results in free carrier-like holons and doublons, leading to disordering dominated by carrier screening and re-ordering dominated by second-order recombination of free holons and free doublons. The distinct non-equilibrium responses of the two correlated states define the richness of behavior accessible by a time-domain view. Overall, the delicate balance or competition between e-e and e-ph interactions may provide an additional angle in understanding and ultimately controlling quantum phases on the moiré landscape.

**Methods**

**60° Device Fabrication and Gating**

The WSe$_2$ and WS$_2$ monolayers are mechanically exfoliated from flux-grown bulk WSe$_2$ crystals and commercially purchased bulk WS$_2$ crystals (HQ Graphene), respectively. Prior to transfer, the crystal orientation of WSe$_2$ and WS$_2$ monolayers are first determined via polarization-resolved second harmonic generation (P-SHG). The monolayers are then stacked together by a dry-transfer technique with a polycarbonate stamp. To distinguish between R- and H-stacked configurations, P-SHG is again applied to the heterobilayer region after the device is fabricated, with results compared to individual monolayers. The error bars for the angle determination are well within ±1°. The sample is grounded via Gr contacts connected to the heterobilayer and single crystal h-BN dielectrics and Gr gates are used to encapsulate the device and provide control of the carrier density and displacement field (if desired) via external source meters (Keithly 2400). SF6 radiofrequency plasma is applied to the stack to etch the encapsulating h-BN and create connection to the Gr gates contacts. All electrodes are defined with electron beam lithography and made of a three-layer metal film of Cr/Pd/Au (3 nm/17 nm/60 nm).

**Spectroscopic Measurements**

For the spectroscopic measurements, the sample is cooled down to T = 11 K under vacuum (<10$^{-6}$ torr) with a closed-cycle liquid helium cryostat (Fusion X-Plane, Montana Instruments). Steady-state reflectance measurements are carried out with a 3200 K halogen lamp (KLS EKE/AL). A 715 nm low pass filter is employed to avoid potential photodoping effects. Following collimation, the lamp light is focused onto the sample/substrate with a 100X, 0.75 NA objective. The reflected light is collected by the same objective and then dispersed with a spectrometer onto an InGaAs array (PyLoN-IR, Princeton Instruments). The steady-state reflectance spectrum is obtained by contrasting the reflected signal from the sample ($R$), which includes the bilayer and dual gate region, and the effective substrate ($R_0$), which includes a region with gates and no bilayer as follows: $(R-R_0)/R_0$. For the spatially-resolved reflectance experiments used for mapping of the bilayer, a dual-axis galvo mirror scanning system is employed.[48] Following spatial scanning of the sample as controlled via the angles of the galvo mirrors, the reflected light is spatially filtered through a pinhole and collected by the detector.

The steady-state photoluminescence (PL) experiments are performed with a HeNe laser (Model 31-2140-000, Coherent). The excitation power is set to the range 50-100 nW and focused onto the sample. The PL signal was spectrally filtered from the laser using a long-pass filter prior to dispersal with a spectrometer and detection (the same detection system described above is employed here).

The pump-probe experiments are seeded by femtosecond pulses (250 kHz, 1.55 eV, 100 fs) generated by a Ti:sapphire oscillator (Mira 900, Coherent) and regenerative amplifier (RegA 9050, Coherent). The output is then split to form the pump and probe arms. For the probe, a fraction of the fundamental is focused into a sapphire crystal to generate a white light continuum which is then spectrally filtered (750±40 nm band pass filter) to cover the lowest energy $WSe_2$ moiré exciton. The pump beam (800 nm, fundamental) is then directed towards a motorized delay stage to control the time delay, $\Delta t$, and passed through an optical chopper to generate pump-on and -off signals. Following this, the pump and probe arms are directed collinearly to the sample through an objective (100X, 0.75 NA). The pump and probe spot diameters are ~2.36 μm and ~0.9 μm, respectively. The same objective is used to collect the reflected light which is spectrally filtered to remove the pump and dispersed onto an InGaAs detector array (PyLoN-IR, Princeton Instruments). The pump-on and -off spectra at varying $\Delta t$ are then used to calculate the transient reflectance signal ($\Delta R/R$) where $\Delta R = R_{on} - R_{off}$ with the subscript indicates either pump-on or -off spectra.

**Assignment of Filling Factors**

The charge density, $n$, in the $WSe_2/WS_2$ heterostructure, controlled by the applied gate voltages ($V_t$ and $V_b$ for the top and bottom gates, respectively), is determined using the parallel-plate capacitor model: $n = \frac{\varepsilon \varepsilon_0 \Delta V_t}{d_t} + \frac{\varepsilon \varepsilon_0 \Delta V_b}{d_b}$, where $\varepsilon \approx 3$ is the out-of-plane dielectric constant of hBN, $\varepsilon_0$ is the permittivity of free space, $\Delta V_i$ is the applied gate voltage relative to the valence/conduction band edge, and $d_i$ is the thickness of the hBN spacer. The moiré density, $n_0$, is determined based on the moiré lattice constant, $a_M$, and is given by $n_0 = \frac{2}{\sqrt{3}a_M^2}$. The filling factor, $v$, estimated as the ratio between the charge doping and moiré densities, is fit based on the experimental gate-dependent steady-state reflectance and photoluminescence measurements (Fig. 1b and Fig. S1b). For the 60° device, the thickness of the top and bottom hBN spacers is determined to be $d_t \approx 36.4$ and $d_b \approx 39.6$ nm, respectively. The twist angle is determined to be ~0.8°.

**Data Analysis and Further Discussion**

In the following we describe the data processing involved in analyzing both the overall transient signal and the disordering/recovery dynamics. The data is first collected as a function of probe energy and pump-probe delay, $\Delta t$, at varying $V_g$ and pump fluence. To generate Fig. S2-S3, pump-probe spectra are collected at a series of $\Delta t$ at varying $V_g$ and pump fluence. At each $\Delta t$ and for each fluence, the $V_g$-dependent spectra are then combined as a function of probe energy versus $V_g$ to generate two-dimensional pseudo-color plots. We take $V_g = V_t = V_b$ unless otherwise specified. The data from these plots are then integrated in the range 1.67~1.69 eV about the maximum of the lowest energy WSe$_2$ moiré sensor, normalized, offset, and multiplied by -1 to generate Fig. 1c-1e.

To analyze the time traces, the signal in a ~0.01 eV window about the maximum of the lowest energy moiré band of WSe$_2$ is averaged (e.g., Fig. 2a-2b, Fig. S7, and Fig. S8a). For Fig. 2a-2b, these traces are then normalized for better comparison of the dynamics specifically. The overall $|\Delta R/R|$ trend (Fig. S5) is analyzed by averaging the rectangular region about the maximum signal along both the probe energy axis (in a ~0.01 eV window about the maximum of the lowest energy moiré band of WSe$_2$) and temporal dimensions (in the $\Delta t$ window from 7.5~8.5 ps).

To isolate the dynamics from the abovementioned time traces, fits are performed. For $v = -1$ and -2, the normalized fits compared directly to the data are shown in Fig. 2a-2b (black lines) as well as separately in Fig. 2c-2d. The corresponding normalized residuals are shown in Fig. S6. A comparison between the unnormalized fits and data for both states is shown in Fig. S7. For $v = 0$, Fig. S8a shows the fits (where applicable) in red where the corresponding residuals are shown in Fig. S8b-S8c. For all fits, the coherent artifact region arising from pump-probe overlap is omitted. All errors are given by 99% confidence intervals.

The $v = -1$ data for pump fluence $f = 3, 7, 10$, and 14 µJ/cm$^2$ are fit by a biexponential function with disordering captured by the functional form $A_d \exp[t/\tau_d]$ and recovery by $A_r \exp[t/\tau_r]$. The $v = -1$ data for pump fluence $f = 28, 56$, and 85 µJ/cm$^2$ are fit using the aforementioned form with the inclusion of an additional term, $A_\infty \exp[t/\tau_\infty]$, to account for the additional component relating to thermionic emission (discussed further below). Here, $\tau_\infty = 500$ ps is fixed as the timescale for this process is beyond the duration of the present experiment and is therefore unresolvable. Fig. 3-4 show the obtained time constants for disordering, $\tau_d$, and recovery, $\tau_r$, as a function of fluence. Fig. S10 shows the corresponding rate of recovery, $k_r = 1/\tau_r$, as a function of fluence.

The $v = -2$ data for pump fluence $f = 3, 7, 10, 14, 28$, and 56 µJ/cm$^2$ are fit using a function with disordering captured by the functional form $A_d \exp[t/\tau_d]$ and recovery by $A_r /(1 + [A_r t/\tau_{r,\text{eff}}])$. As explained in the text and supported by Fig. 4 ($\tau_r/\tau_{r,\text{eff}}$ as a function of fluence) and Fig. S10 ($k_r/k_{r,\text{eff}}$ as a function of fluence), the recovery for $v = -2$ is best represented by a second order process as captured by the employed functional form for recovery. Fig. 3 shows the corresponding $\tau_d$ as a function of fluence. The $v = -2$ data for pump fluence $f = 85$ µJ/cm$^2$ is fit using the above form with the inclusion of an additional term, $A_\infty \exp[t/\tau_\infty]$, to account for the additional component relating to thermionic emission (discussed further below). Here, $\tau_\infty = 500$ ps is fixed as the timescale for this process is beyond the duration of the present experiment. We note that unlike $v = -1$, this term was not necessary for fitting the $f = 28$ and 56 µJ/cm$^2$ data. This is likely because the overall increased signal strength for $v = -2$ versus $v = -1$ obscures contributions from this component at $f = 28$ and 56 µJ/cm$^2$ in the former.

The $v = 0$ data for pump fluence $f = 28, 56$, and 85 µJ/cm$^2$ is fit using a function with rise captured by the functional form $A_\text{rise} \exp[t/\tau_\text{rise}]$ and an additional long-lived component by $A_\infty \exp[t/\tau_\infty]$. Here, $\tau_\infty = 500$ ps is again fixed as the timescale for this process is beyond the duration of the present experiment. The average $\tau_\text{rise} = 1.4 \pm 0.4$ ps (see Fig. S8d). Meaningful fits were unable to be attained for $f < 28$ µJ/cm$^2$.

We attribute the observed dynamics for $v = 0$ with $f > 14$ µJ/cm$^2$ (as well as the presence of the additional long-lived component, $\tau_\infty$, for $v = -1, -2$) to hot hole injection (into the WSe$_2$/WS$_2$ bilayer) via thermionic emission following unavoidable photo-excitation of the Gr electrodes at sufficiently high excitation densities (schematically depicted in Fig. S9)[40,41]. We note that in Fig. S9, the shown Gr and h-BN band alignment is based on Ref. [41], while the WSe$_2$ and WS$_2$ band alignment is based on Ref. [49]. As described in Refs. [40,41], photoexcited carriers in the Gr can rapidly thermalize via scattering (e.g., Auger-like process). Following this, 'hot' carriers (specifically holes) in the tail of the resulting thermal distribution have sufficient energy to overcome the Gr/h-BN/WSe$_2$ (or Gr/h-BN/WS$_2$) energetic barrier (i.e., valence band offset) and to undergo interlayer transfer. Further, this process is highly pump fluence-dependent, as also captured clearly in the time traces of the $v = 0$ where dynamics are only observed for $f > 14$ µJ/cm$^2$. This is consistent with the additional long-lived component only being necessary in this same high fluence regime for proper fitting of the correlated state dynamics. Interlayer charge transfer ultimately driven via thermionic emission is linked to a multi-component timescale with a faster (< 90 fs) and a slower (~1 ps) contribution.[41] While the former is unresolvable within our instrument response function (and is mainly related to the thermalization rather than the transfer process), the latter agrees well with the fit $\tau_{rise}$ for $v = 0$ as expected (and is mainly related to interlayer transfer). The long-lived component ($\tau_\infty$) in our measurement (for $v = 0, -1, -2$) may result from, in addition to this thermionic hole transfer and charging, transient lattice heating. This process does not interfere with the observed disordering/reordering of the $v = -1, -2$ states.

**Acknowledgements**

This work was supported by Programmable Quantum Materials, an Energy Frontier Research Center funded by the U.S. Department of Energy (DOE), Office of Science, Basic Energy Sciences (BES), under award DE-SC0019443. XYZ acknowledges support by DOE-BES under award DE-SC0024343 for the development of the pump-probe method on moiré systems. EAA gratefully acknowledges support from the Simons Foundation as a Junior Fellow in the Simons Society of Fellows (965526). Facilities supported by the Materials Science and Engineering Research Center (MRSEC) through NSF grant DMR-2011738 were utilized in this work. K.W. and T.T. acknowledge support from the JSPS KAKENHI (Grant Numbers 21H05233 and 23H02052) and World Premier International Research Center Initiative (WPI), MEXT, Japan. We thank Yinjie Guo for assistance with sample mounting and Dipti Jasrasaria and Timothy Berkelbach for helpful discussions.

**Author Contributions**

EAA and XYZ conceived this work. EAA and YL carried out all spectroscopic measurements. BY was responsible for sample fabrication and characterization, under the supervision of CRD. TT and KW provided the h-BN crystal. JCH provided the WSe$_2$ crystal. EAA and XYZ interpreted the results, with input from XX. The manuscript was prepared by EAA and XYZ in consultation with all other authors. XYZ supervised the project. All authors read and commented on the manuscript.

**Competing Interests**

The authors declare no competing interests.

**Data Availability Statement**

The data within this paper are available upon reasonable request.

**Extended Data**

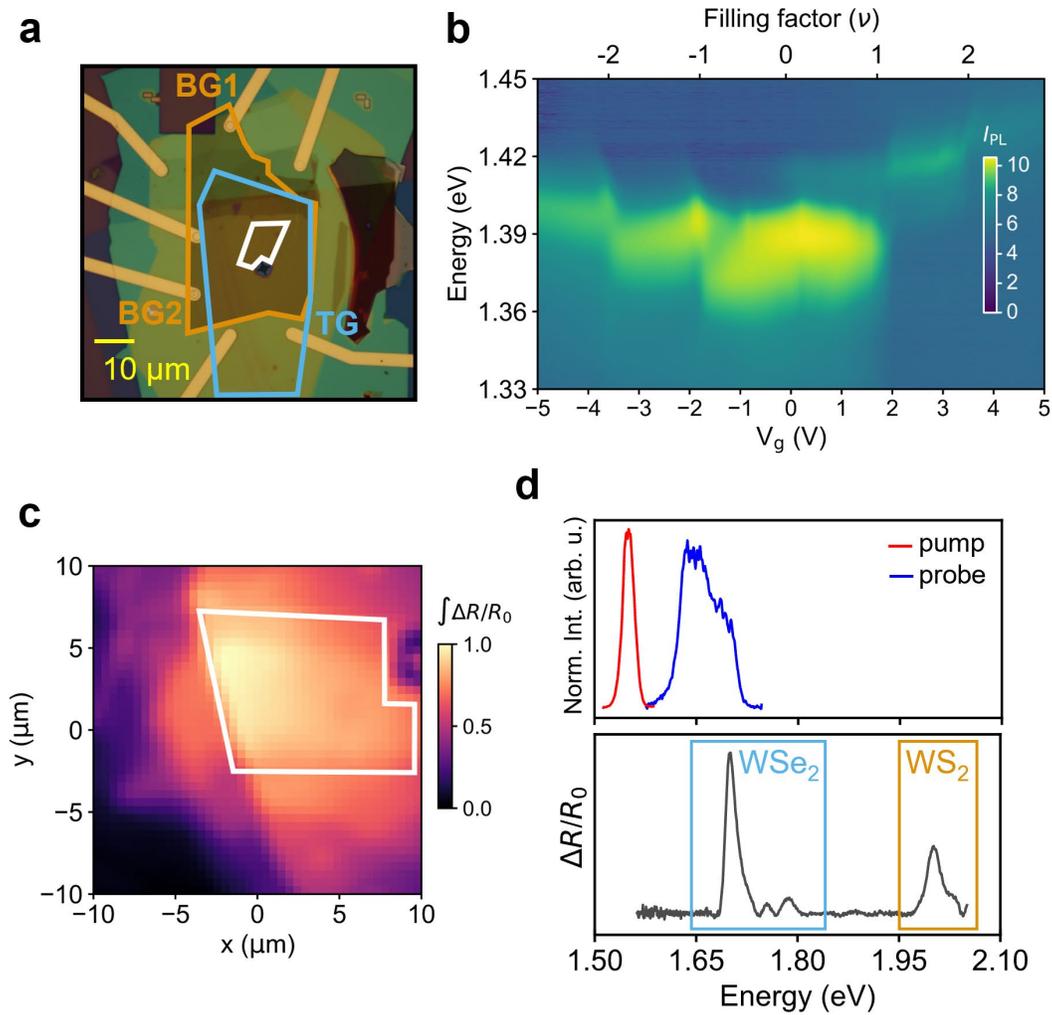

**Figure S1. Additional device characterization. a** Optical image of the device where the top (TG) and bottom (BG1, BG2) gates are labeled, along with outlines showing the Gr electrodes (blue and orange outlines corresponding to the top and bottom electrodes, respectively) and the WSe$_2$/WS$_2$ overlap region (white outline). **b** Gated ($V_g = V_t = V_b$) steady-state photoluminescence spectrum (with intensity shown on a logarithmic scale). **c** Steady-state reflectance contrast mapping with the white line approximately highlighting the WSe$_2$/WS$_2$ overlap region. **d** Shown in the top panel are the normalized pump and probe spectra in red and blue, respectively. Shown in the bottom panel is the steady state reflectance spectrum for the device (undoped). The blue (orange) box highlights the WSe$_2$ (WS$_2$) exciton region. The multiple peak structure in the WSe$_2$ exciton region (rather than the observation of the single WSe$_2$ A exciton) indicates the formation of moiré excitons, highlighting the quality of the moiré interface. All data were collected at 11 K. Above, $\Delta R = R - R_0$ where $R$ is the reflected signal from the bilayer and dual gate region while $R_0$ is the reflected signal from a region with gates and no bilayer.

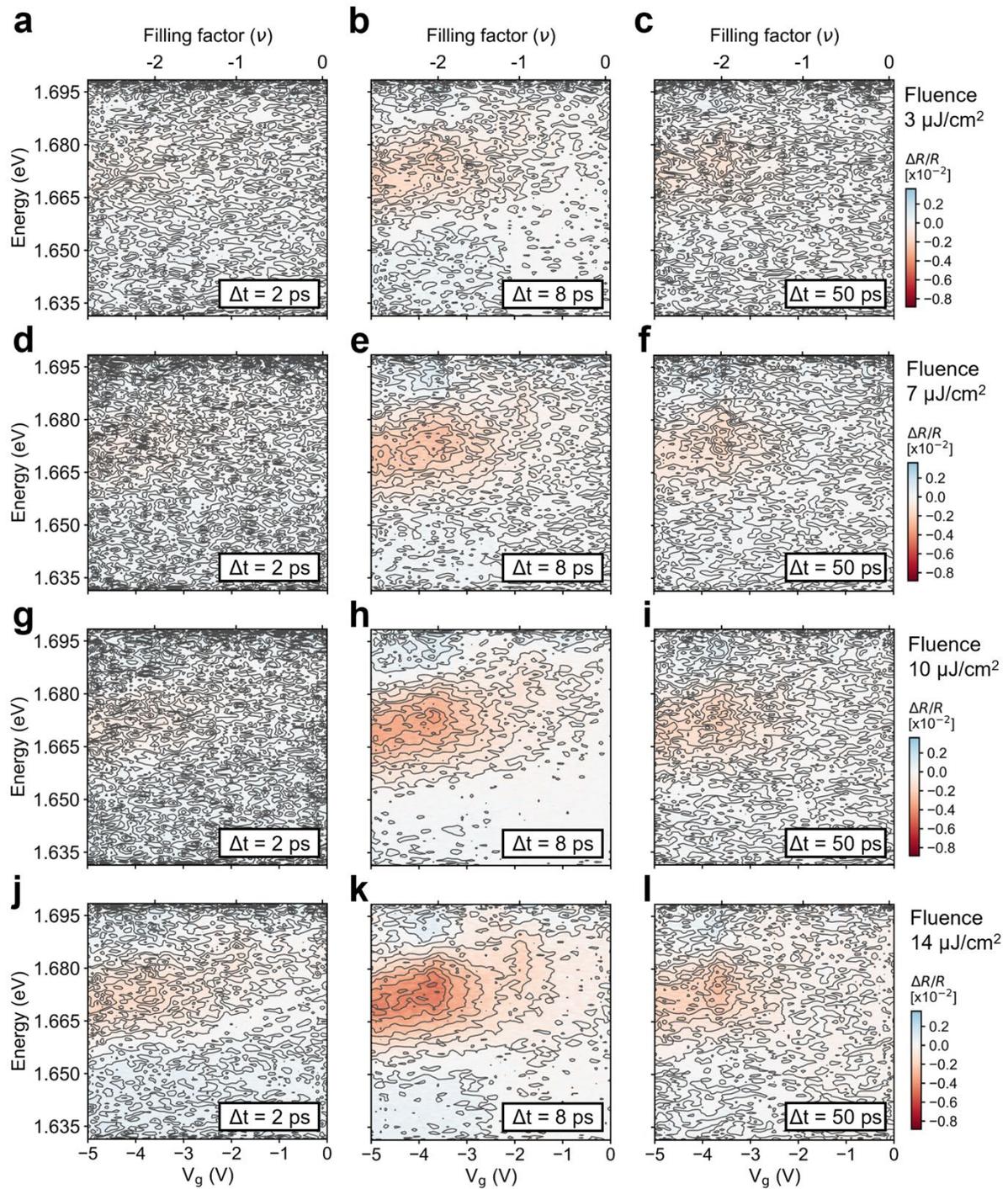

**Figure S2. Fluence-dependent transient $V_g$ mapping (part I).** Transient reflectance spectra for the device plotted as a function of probe energy (covering the lowest energy moiré band of $WSe_2$) versus gate voltage ($V_g = V_t = V_b$) and filling factor at specific pump-probe delays ($\Delta t$ = 2, 8, 50 ps in columns one through three, respectively). The data were collected at 11 K. The colormaps have been scaled to those in Fig. S3d-f to allow for a meaningful overall intensity comparison as a function of fluence within/between Fig. S2-S3. The employed pump fluences are as follows: 3 μJ/cm² (panels a-c), 7 μJ/cm² (panels d-f), 10 μJ/cm² (panels g-i), and 14 μJ/cm² (panels j-l).

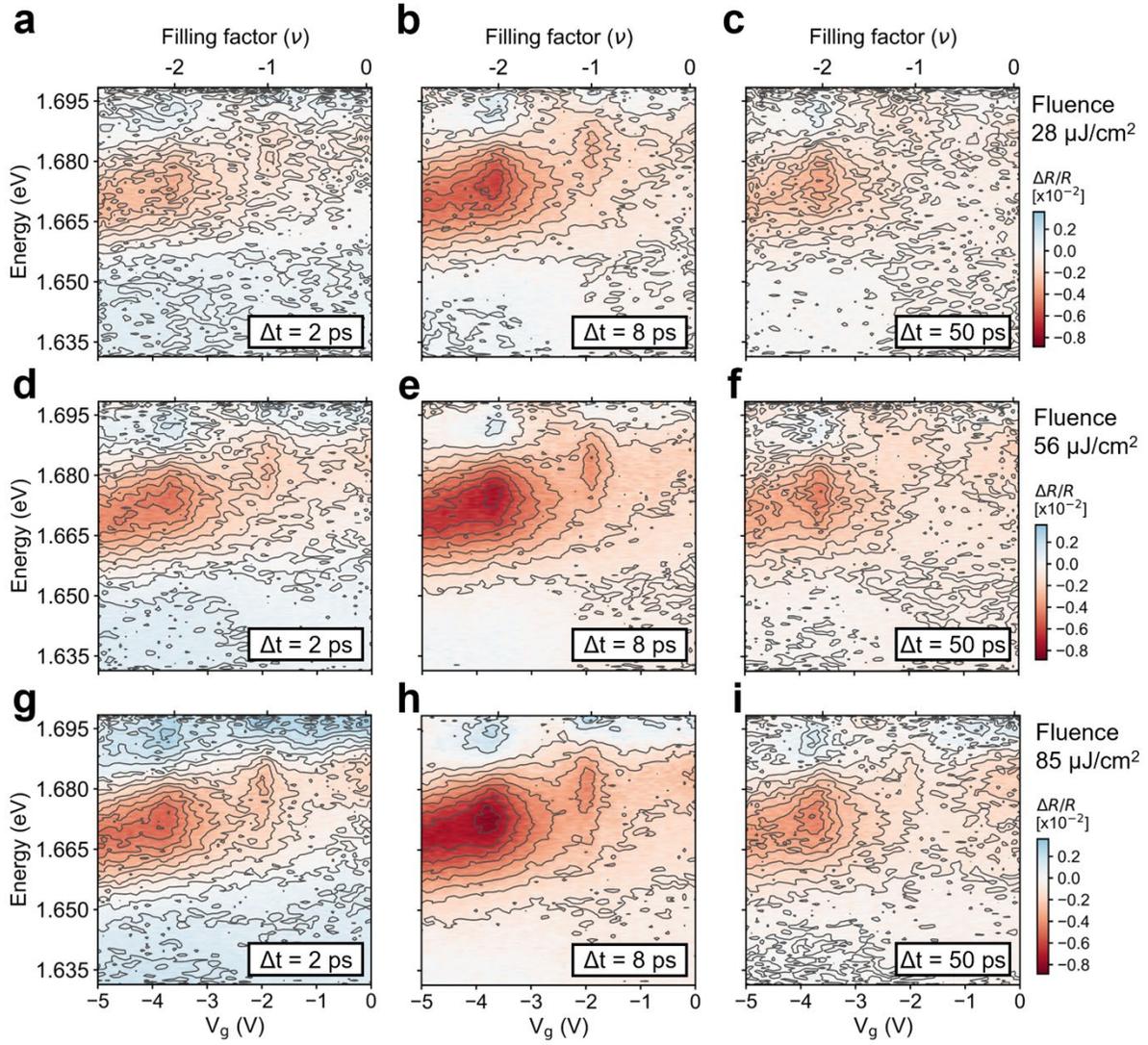

**Figure S3. Fluence-dependent transient $V_g$ mapping (part II).** Transient reflectance spectra for the device plotted as a function of probe energy (covering the lowest energy moiré band of $WSe_2$) versus gate voltage ($V_g = V_t = V_b$) and filling factor at specific pump-probe delays ($\Delta t$ = 2, 8, 50 ps in columns one through three, respectively). The data were collected at 11 K. The colormaps have been scaled to those in Fig. S3d-f to allow for a meaningful overall intensity comparison as a function of fluence within/between Fig. S2-S3. The employed pump fluences are as follows: 28 µJ/cm² (panels a-c), 56 µJ/cm² (panels d-f), and 85 µJ/cm² (panels g-i).

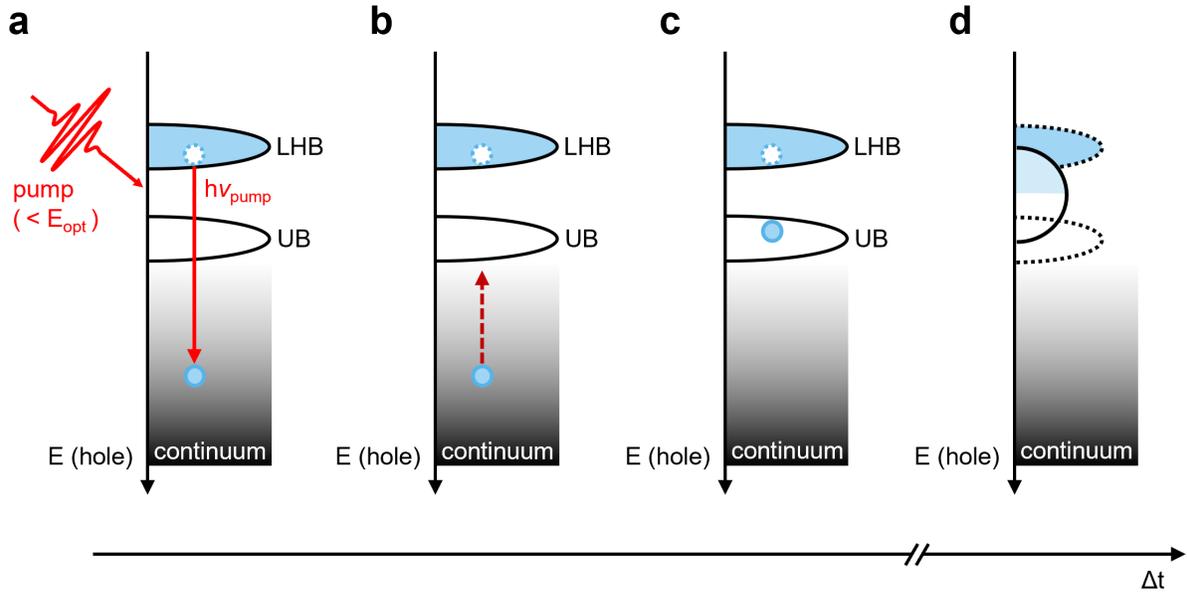

**Figure S4. Schematic of pump-induced disordering. a** As described in the text, absorption of a pump photon (with energy < $E_{opt}$, red arrow) excites a hole (blue) from the lower Hubbard band (LHB) deep into the valences bands(s) (gray shaded region). **b** The photoexcited hole (blue sphere) relaxes within this continuous manifold (dark red arrow) on ultrafast timescales towards the upper unoccupied band(s) (UB, the upper Hubbard band or charge transfer band). **c** This leads to the formation of holons and doublons (white and blue spheres, respectively) across the correlation gap. **d** The presence of holons and doublons leads to disordering of the correlated insulator state (on some characteristic timescale) and (partial) gap closing. We note that the hole energy scale is opposite to that of an electron scale. The time axis below indicates the temporal ordering of these processes.

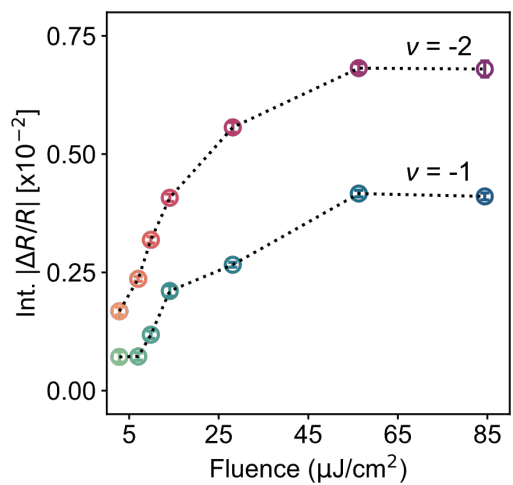

**Figure S5. Maximum integrated signal as a function of fluence.** Overall integrated fluence-dependent transient $|\Delta R/R|$ for the $v = -1$ and -2. For each point, the corresponding transient reflectance spectra was averaged in a rectangular region over a ~0.01 eV probe energy range about the maximum of the lowest energy moiré band of $WSe_2$ and over a pump-probe delay ($\Delta t$) window from 7.5~8.5 ps where the signal is at the maximum. The dotted black lines show the linear interpolation between data points to highlight the data trend. Error bars indicate 99% confidence intervals. Large circle outlines have been added to aid with data visualization. The fluence-dependent shaded colors for the $v = -1, -2$ states are consistent with the data shown elsewhere in the text.

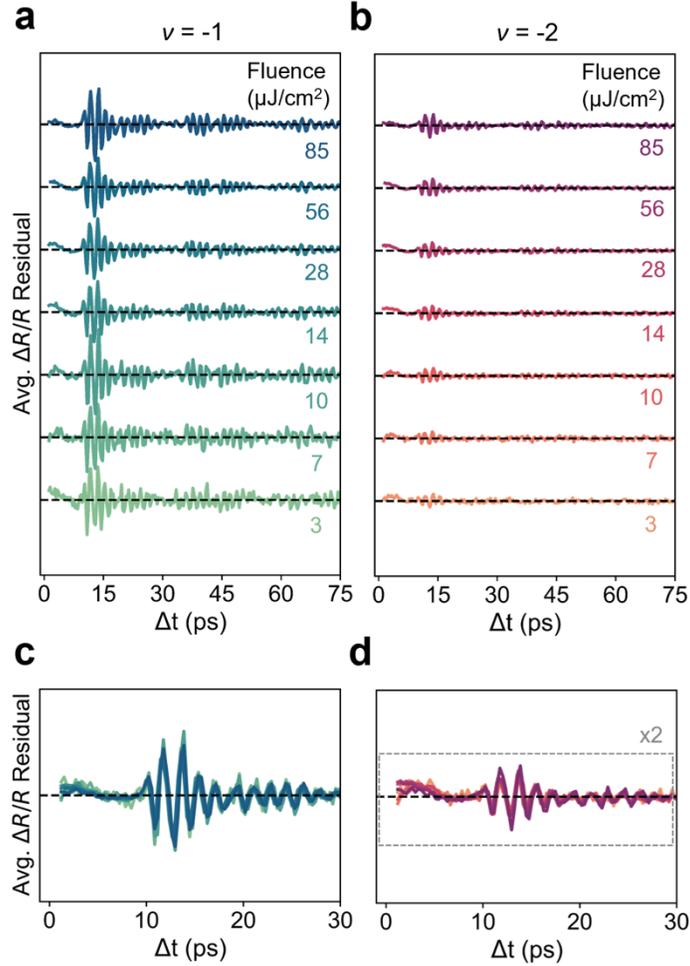

**Figure S6. Fit residuals for the $v = -1$ and $v = -2$ states. a** Normalized fit residuals for the $v = -1$ state as a function of fluence corresponding directly to Fig. 2a. **b** Normalized fit residuals for the $v = -2$ state as a function of fluence corresponding directly to Fig. 2b. **c** Expanded view of the early $\Delta t$ normalized fit residuals for the $v = -1$ state as a function of fluence. **d** Expanded view of the early $\Delta t$ normalized fit residuals for the $v = -2$ state as a function of fluence where the data have been multiplied by a factor of two for clarity. See the Methods section for further details and the employed fit functions. Throughout the panels, increasing pump fluence is shown from light to dark shades. All fluence-dependent data are colored consistently for each of the filling factors throughout the figure and elsewhere in the text. All data were collected at 11 K. In panels c and d, the similarity between the $\Delta t < 10$ ps residuals for both sates as well as a function of fluence is suggestive of a low frequency modulation that may point to the direct observation of the amplitude mode of the charge ordered states. However, the onset of the interlayer phonons obscures potential subsequent oscillation cycles. Further suggestive, this behavior is noticeably absent in the behavior of the control experiments on the $v = 0$ state (Fig. S8).

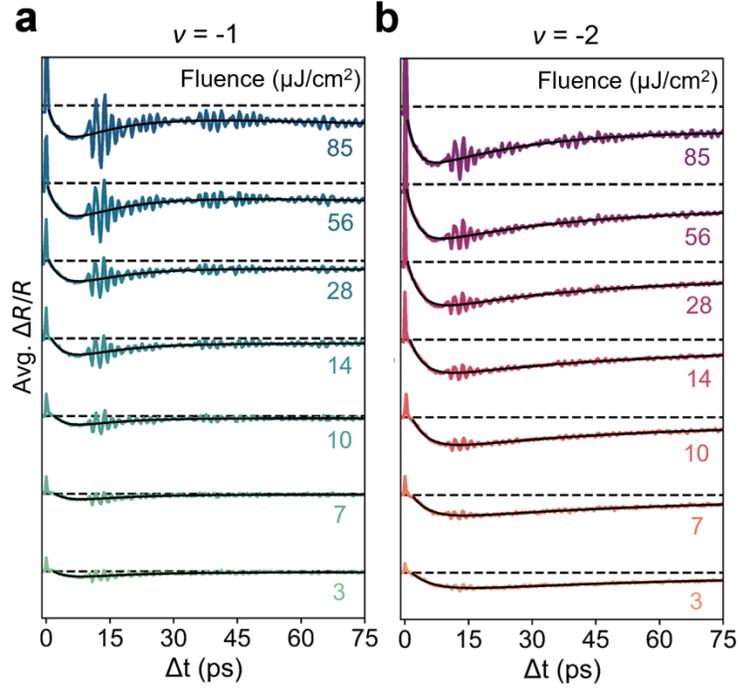

**Figure S7. Unnormalized transient response of $v = -1$ and $v = -2$. a-b** Unnormalized fluence-dependent temporal response of the $v = -1, -2$ states corresponding to Fig. 2a-b, respectively. The time traces were averaged in a narrow window about the maximum of the lowest energy $WSe_2$ moiré exciton sensor. Data are shown in shaded colors while the fits are shown in black. See the Methods section for further fit details. Throughout the panels, increasing pump fluence is shown from light to dark shades. The fluence-dependent shaded colors for the $v = -1, -2$ states are consistent with the data shown elsewhere in the text. All data were collected at 11 K.

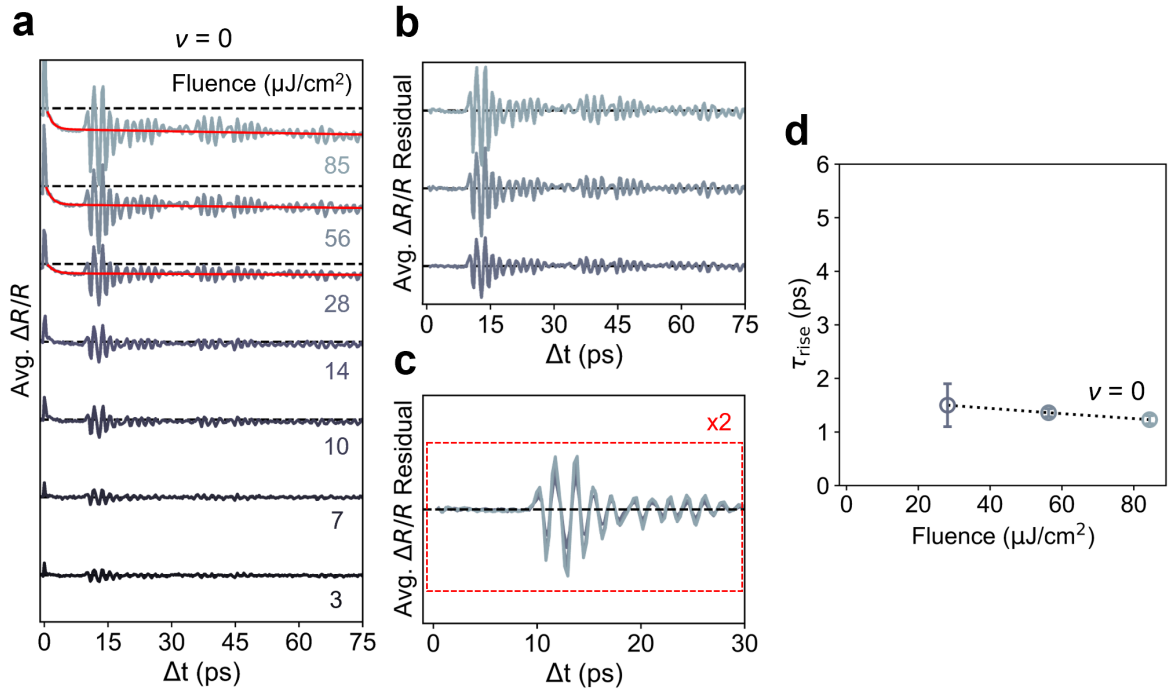

**Figure S8. Transient response of undoped $v = 0$ control. a** Fluence-dependent temporal response of the undoped $v = 0$ state. The time traces were averaged in a narrow window about the maximum of the lowest energy $WSe_2$ moiré exciton sensor. Data are shown in shades of gray (where lighter shades correspond to higher pump fluence, $f$) while the fits used for time constant extraction are shown in red. Only the higher fluence data ($f$ = 28, 56, 85 μJ/cm$^2$) were able to be fit. The fitting form consists of a rapid exponential rise followed by a long-lived rise beyond the duration of the experiment (>>100 ps). **b** Fit residuals corresponding with panel a. **c** Expanded view of the early Δt normalized fit residuals for the $v = 0$ state as a function of fluence for direct comparison with Fig. S6. **d** Initial rise time constant ($\tau_{rise}$) for $v = 0$ as a function of fluence. The dotted black line shows the linear interpolation between data points to highlight the data trend. All error bars indicate 99% confidence intervals. Large circle outlines have been added to aid with data visualization. See the Methods section for further details.

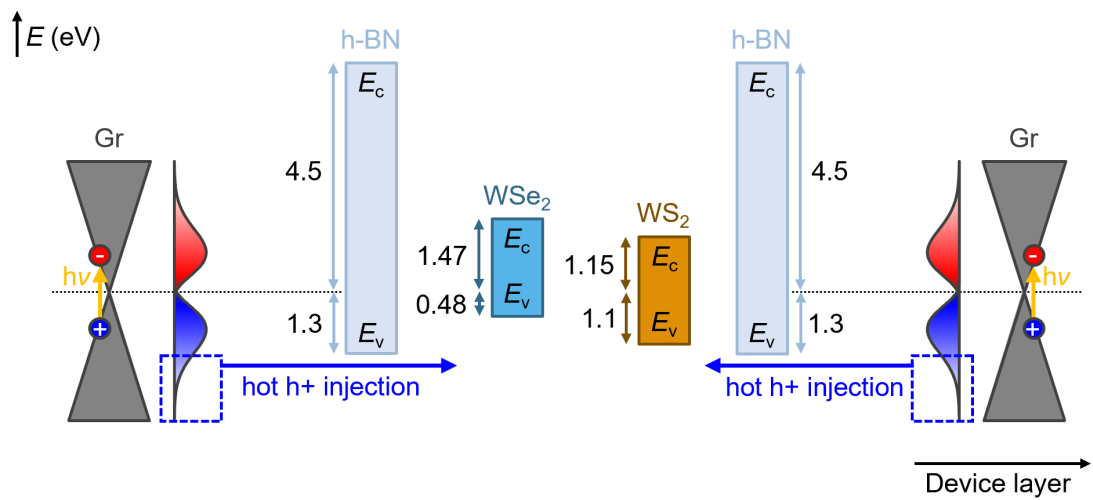

**Figure S9. Schematic of hot hole injection via thermionic emission.** Pump excitation (yellow arrow, with photon energy 1.55 eV) leads to photoexcitation of carriers (electrons and holes shown in red and blue, respectively) in the Gr electrodes. Scattering-induced thermalization leads to thermal electron and hole distributions (shown in red and blue, respectively, next to Gr) with hot carrier tails. Following this, the hot holes (highlighted with dashed blue boxes) have sufficient energy to overcome the energetic barrier to directly undergo interlayer transfer towards the bilayer region (blue arrows). All shown energies are in units of eV. The Gr electrodes, h-BN dielectric spacers, $WSe_2$, and $WS_2$ layers are shown in gray, light blue, blue, and orange, respectively, where the colors are chosen to be consistent with the device structure shown in Fig. 1a. The Gr and h-BN on both the left and right sides of the figure signify the dual-gate structure. See the Methods section for further details.

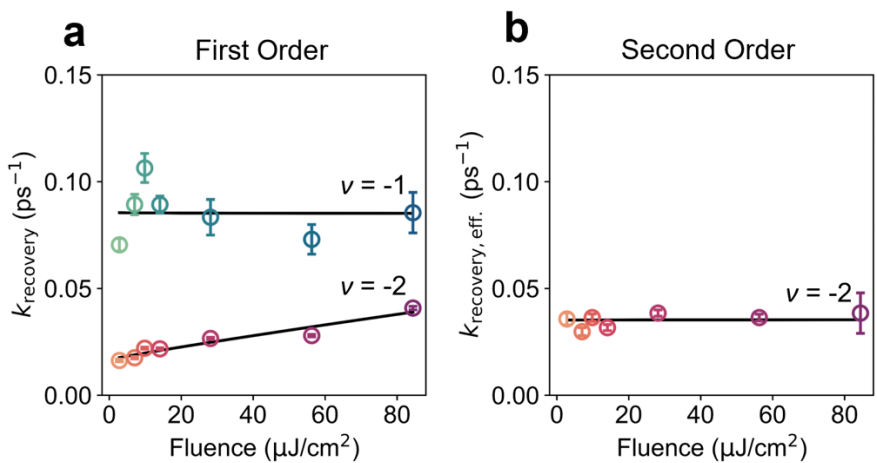

**Figure S10. Recovery rates for $v = -1$ and $v = -2$. a** Recovery rate constant ($k_r = 1/\tau_r$) for $v = -1$ and $v = -2$ as a function of fluence based on a first order recovery process (i.e., first order recovery rate). The black lines show the linear fit trend lines for the fluence dependence of the rate constant, $k_r$. For $v = -2$, $k_r$ was found to be proportional to $n_{ex}^{0.9\pm0.4}$. **b** Effective recovery rate constant ($k_{r,\,eff}$) for $v = -2$ as a function of fluence. Here, $k_{r,\,eff}$ was extracted based on considering the recovery for $v = -2$ as a second order process. The black line shows the linear fit trend line for the fluence dependence of the effective rate constant, $k_{r,\,eff}$. Throughout the figure, all error bars indicate 99% confidence intervals. Large circle outlines have been added to aid with data visualization. The fluence-dependent shaded colors for the $v = -1, -2$ states are consistent with the data shown elsewhere in the text.